\def\RR{{\buildrel {\scriptstyle \hphantom{i}o}\over R}\vphantom{R}}
\def\GG{{\buildrel {\scriptstyle \hphantom{i}o}\over G}\vphantom{G}}
\def\gg{{\buildrel {\scriptstyle \hphantom{i}o}\over g}\vphantom{G}}

\def\a{\alpha}                    
\def\m{\mu}              \def\n{\nu}           \def\k{\kappa}
\def\G{\Gamma}           \def\g{\gamma}        \def\ve{\varepsilon}
           \def\d{\delta}        
           \def\s{\sigma}        
\def\O{\Omega}           \def\o{\omega}        \def\vphi{\varphi}
          \def\l{\lambda}

\documentstyle[preprint,aps,eqsecnum]{revtex}

        \def\nn{\nonumber}
\def\be{\begin{equation}}             \def\ee{\end{equation}}
\def\bea{\begin{eqnarray} }           \def\eea{\end{eqnarray} }
\def\ba#1{\begin{array}{#1}}          \def\ea{\end{array}}
\def\lab#1{\label{#1}}                \def\eq#1{(\ref{#1})}
\def\bsubeq{\begin{mathletters}}      \def\esubeq{\end{mathletters}}
\def\bitem{\begin{itemize}}           \def\eitem{\end{itemize}}

\def\mat#1#2{ \left( \begin{array}{#1} #2 \end{array} \right) }
\begin{document}
\tighten
\title{Consistency analysis of Kaluza-Klein geometric sigma models}
\author{M. Vasili\'c\thanks{E-mail address: mvasilic@phy.bg.ac.yu}}
\address{Institute of Physics, P.O.Box 57, 11001 Beograd, Yugoslavia}
\date{\today}
\maketitle
\begin{abstract}
Geometric $\s$-models are purely geometric theories of scalar fields
coupled to gravity. Geometrically, these scalars represent the very
coordinates of space-time, and, as such, can be gauged away. A
particular theory is built over a given metric field configuration
which becomes the vacuum of the theory. Kaluza-Klein theories of the
kind have been shown to be free of the classical cosmological
constant problem, and to give massless gauge fields after
dimensional reduction. In this paper, the consistency of dimensional
reduction, as well as the stability of the internal excitations, are
analyzed. Choosing the internal space in the form of a group
manifold, one meets no inconsistencies in the dimensional reduction
procedure. As an example, the $SO(n)$ groups are analyzed, with the
result that the mass matrix of the internal excitations necessarily
possesses negative modes. In the case of coset spaces, the
consistency of dimensional reduction rules out all but the stable
mode, although the full vacuum stability remains an open problem.
\end{abstract}

\narrowtext
\section{Introduction} 

Geometric $\s$-models have originally been proposed as an attempt to
explain the pure geometric origin of fermionic matter. Indeed, it
has been shown in \cite{sigma} that scalar matter can be coupled to
gravity in such a way that two goals are achieved. First, the theory
possesses a kink solution with topologically nontrivial scalar
sector which allows for the fermionic type of quantization. Second,
using diffeomorphism invariance, all the scalar fields can be gauged
away giving the theory a purely geometric form. The possibility of
geometrizing fermionic matter has a solid basis in the early work of
Finkelstein and Rubinstein \cite{Finkelstein} who realized the role
of multiply-connected configuration spaces for the existence of
fermions. An example of the kind is the configuration space of
four-dimensional gravitational kinks of Finkelstein and Misner
\cite{Misner}. Its double-connectedness enables the existence of
double-valued wave functions but, unfortunately, it is in no way
related to the spin of the system. The Ref. \cite{sigma},
however, uses as its role model the 't Hooft-Polyakov monopole
solution of the $SO(3)$ gauge theory spontaneously broken to $U(1)$
by a Higgs triplet \cite{Olive}. It has been shown \cite{Ringwood}
that these monopoles admit both half-integer spin and fermion
statistics in the sense of Finkelstein and Rubinstein. The necessary
multiple connectedness of the configuration space, however, stems
from the Higgs triplet exclusively, and is not directly related to
the gauge fields. Using this idea, the same goal has been achieved
in \cite{sigma} by identifying the coordinates of space-time with the
components of a set of scalar fields. The resulting theory has a
form of a nonlinear $\s$-model coupled to gravity, and necessarily
possesses a solution very much similar to the 't Hooft-Polyakov
monopole. The difference is that the scalar fields of this solution
have a pure geometric meaning---they are just the coordinates of our
space-time, and can be gauged away.

The actual procedure of constructing geometric $\s$-models begins by
specifying a fixed metric field configuration $\gg_{\mu\nu}$ which
later becomes the vacuum of the theory. The dynamics is chosen from
a variety of possibilities. The simplest one is given by the
Einstein like equations of the form $R_{\mu\nu}=\RR_{\mu\nu}$, where
the fixed function $\RR_{\mu\nu}$ stands for the Ricci tensor of the
vacuum metric $\gg_{\mu\nu}$. These are the non-covariant field
equations whose non-vanishing right-hand side actually {\it defines}
matter, and which, by construction, possess the classical solution
$g_{\mu\nu}=\gg_{\mu\nu}$. The covariantization of the theory is
achieved by employing a new set of coordinates, say $\O^i=\O^i(x)$,
to fix the Ricci tensor on the right-hand side. Then, the equations
of motion take the form of a nonlinear $\s$-model coupled to
gravity, with the scalar sector consisting of as many scalar fields
$\O^i(x)$ as the number of space-time dimensions. By choosing the
gauge $\O^i=x^i$, the field equations are brought back to their
non-covariant but purely geometric form $R_{\mu\nu}=\RR_{\mu\nu}$.
The multiple-connectedness of the configuration space is a
consequence of the topologically nontrivial one-to-one mapping
$\O^i=x^i$.

The idea of geometric $\s$-models has further been developed in
\cite{KKsigma}, and applied to Kaluza-Klein theories. Using the fact
that geometric $\s$-models are built over freely chosen ground
states, one can build a Kaluza-Klein theory of the kind by
specifying the vacuum geometry in the form of the direct product of
the 4-dimensional Minkowski space $M^4$ with the internal
$d$-dimensional space $B^d$. The resulting theory will necessarily
posses the classical solution $M^4\times B^d$, and, therefore, be
free of the classical cosmological constant problem. An action
functional of this kind has already been discussed in literature.
The authors of Refs. \cite{Omero} and \cite{Gell-Mann} have employed
scalar fields in the form of a nonlinear $\s$-model to trigger the
compactification, but failed to obtain massless gauge fields. In
\cite{KKsigma}, however, this problem has been solved by abandoning
the simple dynamics given by $R_{MN}=\RR_{MN}$ in favor of
$R^{MN}=\RR^{MN}$, the functions on the right-hand side standing for
the Ricci tensor of the Kaluza-Klein vacu\-um metric $\GG_{MN}$ of
the needed form $M^4\times B^d$. Owing to the non-covariant form of
the above equations, the two respective theories are inequivalent.
When modified by adding terms proportional to $(G_{MN}-\GG_{MN})$,
the equations of motion allow for the construction of a simple
Lagrangian. It is this Lagrangian which will be the subject of our
analysis in the subsequent sections.

The lay-out of the paper is as follows. In Section II, we shall
define our model, and analyze its symmetry properties. The theory
turns out to have a gauge symmetry bigger than pure diffeomorphism
invariance. Owing to this, we shall be able to demonstrate how the
gauge fixing of the complete scalar sector still leaves us with the
standard 4-dimensional gauge invariance. The basic results of
Ref. \cite{KKsigma} are then recollected with the emphasis on
the unsolved stability problem. In Section III, we shall study some
consistency aspects of the dimensional reduction procedure. In
particular, the Lagrangian constraints of the theory are recognized
in the gauge fixed matter field equations, and their response to the
dimensional reduction ansatz is analyzed. In the case of $B^d=S^2$,
we shall find that all five unstable modes \cite{KKsigma} are ruled
out by the consistency requirements. On the other hand, choosing
$B^d$ in the form of a group manifold is shown to be consistent with
the complete set of equations of motion. In Section IV, we shall
calculate the mass matrix of the internal excitations of the $SO(n)$
group manifolds, with the result that all of them contain negative
modes.

Section V is devoted to concluding remarks.

\section{Lagrangian and symmetries}

The model we are going to explore consists of Einstein gravity in
$4+d$ dimensions coupled to $4+d$ scalar fields $\O^A(X)$, as given
by the following action functional:
\be
I = - \k^2\int d^{4+d}X\sqrt{-G}\left[ R + F^{AB}(\Omega)
{{\partial X^M}\over{\partial \Omega^A}}{{\partial X^N}
\over{\partial \Omega^B}}G_{MN} - V(\Omega) \right] .
                                                            \lab{2.1}
\ee
The target metric $F^{AB}(\Omega)$ and the potential $V(\Omega)$ are
defined through
\[
F^{AB}(\Omega)\equiv \RR^{AB}(\Omega)  \, ,\qquad
V(\Omega)\equiv 2\RR(\Omega) \, ,
\]
where $\RR^{MN}(X)$ and $\RR(X)$ stand for the vacuum values of the
Ricci tensor and scalar curvature, respectively. As explained in the
introduction, the vacuum metric $\GG_{MN}$ is fixed in advance, and
we choose it to be the direct product of the 4-dimensional Minkowski
space $M^4$ and the internal $d$-dimensional compact space $B^d$:
\be
\GG_{MN}\equiv \mat{cc}{\eta_{\mu\nu}   &0  \\
                             0     &\phi_{mn}(y) }  ,   \lab{2.2}
\ee
Here, $\phi_{mn}(y)$ stands for the metric of $B^d$, and the
coordinates $X^M \equiv (x^{\mu}, y^m)$ are decomposed into
4-dimensional $x^{\mu}$, and internal $y^m$.

The theory given by \eq{2.1} differs from the conventional
$\s$-models by employing the inverse of $\O^A_{\ ,\,M}$ rather than
$\O^A_{\ ,\,M}$ itself. We shall see, however, that this
non-polynomial dependence on the scalar field derivatives is a pure
gauge, and can easily be removed. This is why we postpone the
inspection of the full set of field equations until the symmetry
analysis is done. Let us only notice that
\be
G_{MN}=\GG_{MN} \, ,\qquad \O^A=X^A                       \lab{2.3}
\ee
is easily checked to be a solution of the equations of motion. It
illustrates the geometric origin of the scalar fields $\O^A(X)$, and
is a consequence of the construction procedure described in
\cite{KKsigma}.

The covariant form of the action functional \eq{2.1} tells us that
our theory is invariant under general coordinate transformations
$X^M \rightarrow X^M + \xi^M(X)$. The scalar fields $\O^A$ and the
metric $G_{MN}$ transform in the usual way:
\be
\O^{A^{\prime}}= \O^A - \xi^M \O^A_{\ ,\,M}\ ,
\qquad
G^{\prime}_{MN}= G_{MN}-\xi^L_{\,,\,M}\,G_{LN}-
\xi^L_{\,,\,N}\,G_{LM}-\xi^L G_{MN,\,L} \ ,                \lab{2.4}
\ee
where $\xi^M(X)$ are arbitrary functions of all $4+d$ coordinates.
The full symmetry of the action is, however, not exhausted by the
general coordinate transformations. Owing to our special choice of
the vacuum metric $\GG_{MN}$, the corresponding Ricci tensor
$\RR^{MN}$ is block diagonal with $\RR^{mn}(y)$ the only non-zero
components. This means that our target metric $F^{AB}(\O)$ and the
potential $V(\O)$ are independent of $\O^{\m}$, and that only
$F^{mn}$ components survive in \eq{2.1}. If, in addition, we choose
our internal space $B^d$ to be symmetric, with $m$ Killing vectors
$K^l_a(y)$, $a=1,...,m\,$, the action functional \eq{2.1} will have
an extra internal symmetry of the form
\be
\O^{{\m}^{\prime}}= \O^{\m} + \epsilon^{\m}(\O^{\n})\ ,
\qquad
\O^{m^{\prime}}=\O^m + \epsilon^a(\O^{\n})\,K_a^m(\O^n)   \lab{2.5}
\ee
where $\epsilon^{\m}$ and $\epsilon^a$ are arbitrary functions of
$\O^{\m}$. When applied to small excitations of the vacuum, let us
say
\[
\O^A(X) \equiv X^A + \o^A(X) \ ,
\]
the scalar part of the transformation laws \eq{2.4} and \eq{2.5}
takes the form
\be
\o^{\m^{\prime}} = \o^{\m} + \epsilon^{\m}(x) -
                   \xi^{\m}(x,y) \, ,           \qquad
\o^{m^{\prime}} = \o^m + \epsilon^a(x)\,K_a^m(y) -
                  \xi^m(x,y)\, .                          \lab{2.6}
\ee
We see that it is possible to fix the gauge $\o^A=0$, or
equivalently
\be
\O^A(X) = X^A \ ,                                         \lab{2.7}
\ee
thereby reducing the action \eq{2.1} to a non-covariant but pure
metric form
\be
I = - \k^2\int d^{4+d}X\sqrt{-G}\left[ R-\RR+\RR^{MN}
      \left( G_{MN}-\GG_{MN} \right)\right] .
                                                          \lab{2.8}
\ee
The gauge condition \eq{2.7} constrains our gauge parameters to
satisfy
\be
\xi^{\m}(x,y) = \epsilon^{\m}(x) \, ,\qquad
\xi^m(x,y) = \epsilon^a(x)\, K_a^m(y) \, .                \lab{2.9}
\ee
Therefore, the gauge fixed theory \eq{2.8} is still invariant under
the restricted general coordinate transformations, as given by
\eq{2.9}. Notice that this is exactly the form of symmetry obtained
in the standard Kaluza-Klein treatments. Owing to this, the
effective 4-dimensional theory will have the well known structure of
a non-Abelian Yang-Mills theory coupled to gravity.

Let us now recollect the basic results of Ref. \cite{KKsigma}.
The equations of motion obtained by varying the gauge fixed action
\eq{2.8} have the form
\be
R^{MN} = \RR^{MN} - {2\over{2+d}}\,G^{MN}\RR^{LR}
         \left( G_{LR} - \GG_{LR} \right) ,              \lab{2.10}
\ee
and {\it coincide with the gauge fixed equations of motion of the
covariant action} \eq{2.1}. Indeed, the matter field equations of
\eq{2.1} boil down to
\be
\left(\RR^{LM}G_{LA}\right)_{\!\!,\,M} +
\G^N_{\ NM}\, \RR^{LM}G_{LA} +
{1\over 2}\,\RR^{LM}_{\ \ \ ,\,A}\,G_{LM} -
\RR_{\,,\,A} = 0
                                                         \lab{2.11}
\ee
when the gauge condition \eq{2.7} is imposed, and are easily shown
not to be independent equations of motion. Instead, they follow from
the Bianchi identities applied to \eq{2.10}, and represent the {\it
standard constraint equations} of generally covariant theories. Now,
we see that the vacuum metric $\GG_{MN}$ is an obvious solution to
the equations of motion. When chosen in the form \eq{2.2}, it gives
the Kaluza-Klein theory free of the classical cosmological constant
problem.

The dimensional reduction procedure begins with the standard $4+d$
decomposition \cite{Salam}
\be
G_{MN}\equiv \mat{cc}{g_{\mu\nu}+B^k_{\mu}B^l_{\nu}u_{kl}
&B^k_{\mu}u_{kn}\\ B^k_{\nu}u_{km}&u_{mn}  } ,           \lab{2.12}
\ee
where $g_{\m\n}$ and $u_{mn}$ are further decomposed as
\be
g_{\mu\nu}=\eta_{\mu\nu}+h_{\mu\nu} \, ,\qquad
u_{mn}=\phi_{mn}+\varphi_{mn}\, .                        \lab{2.13}
\ee
The internal manifold $B^d$ is supposed to be a homogeneous space
with $m$ Killing vectors $K_a^l(y)$ which form a (generally
overcomplete) basis in $B^d$. By projecting the metric components on
this basis, let us say
\be
B^m_{\nu} = K^m_aA^a_{\nu} \ ,  \qquad
\varphi_{mn} = K_{am}K_{bn}\,\varphi^{ab}  \, ,          \lab{2.14}
\ee
we obtain the set of field variables suitable for dimensional
reduction. The Latin indices $m$,$n$,... are raised and lowered by
the internal vacuum metric $\phi_{mn}$. The dimensional reduction
ansatz is defined through the constraints
\be
g_{\m\n} = g_{\m\n}(x)\ , \qquad
A^a_{\m} = A^a_{\m}(x)\ , \qquad
\vphi^{ab} = \vphi^{ab}(x)\, .                           \lab{2.15}
\ee
This ansatz is applied to the linearized field equations \eq{2.10}
which are then averaged over the internal coordinates. To simplify
the analysis, we choose our internal space to be an Einstein
manifold
\[
\RR_{mn} = \l\, \phi_{mn}
\]
with $\l<0$ in accordance with the adopted conventions ($\ R^M_
{\ NLR}=\G^M_{\ NL,R}-\cdots \ $, $diag(G_{MN})=(-,+,...,+)\ $).
Then, the effective 4-dimensional equations become
\bsubeq\lab{2.16}
\be
{\cal R}_{\mu\nu}+{1\over 2}\,\vphi_{,\,\mu\nu}
+{{2\l}\over{d+2}}\,\eta_{\m\n}\,\vphi = O(2) \, ,     \lab{2.16a}
\ee
\be
\g_{ab}\,\partial_{\nu} F^{b\mu\nu} = O(2)  \, ,       \lab{2.16b}
\ee
\be
{\s}_{abcd}\,\Box\,\vphi^{cd} +
\mu_{abcd}\,\vphi^{cd} = O(2) \, .                     \lab{2.16c}
\ee
\esubeq
Here, ${\cal R}_{\mu\nu}$ is the Ricci tensor of the metric
$g_{\m\n}$, $\Box$ is the corresponding d'Alembertian, and
$F^a_{\mu\nu}\equiv A^a_{\mu ,\,\nu}-A^a_{\nu ,\,\mu}+O(2)$ is
the gauge field strength for the gauge fields $A^a_{\mu}$. The
coefficients in \eq{2.16} are expectation values of products of the
Killing vectors and their covariant derivatives, as explicitly shown
in \cite{KKsigma}. In particular,
\[
{\g}_{ab}\equiv \langle K^m_a K_{bm}\rangle
\]
is used to raise and lower the group indices. The scalar field
$\vphi\equiv\vphi^a_a$ is shown to satisfy
\be
\left(\Box - {{8\l}\over{d+2}}\right)\vphi = O(2) \, .  \lab{2.17}
\ee
independently of the choice of the Einstein manifold $B^d$. We see
that the conventional choice $\l<0$ ensures the correct sign for its
mass term. As for the first of the equations \eq{2.16}, it is easily
brought to the standard Einstein form by the rescaling
\[
g_{\m\n}\rightarrow \left( 1+{{\vphi}\over 2}\right)
g_{\m\n}+O(2) \, .
\]
The classical linear stability of our effective theory rests upon
the signature of the mass matrix $\mu_{abcd}\,$. Unlike their trace
mode, the traceless components of the scalar excitations
$\vphi^{ab}$ have masses which depend on the particular choice of
$B^d$. In the case of $B^d=S^2$, for example, it has been shown in
\cite{KKsigma} that all five traceless modes have the same negative
mass square equaling ${{4\l}\over 7}\,$. In the subsequent sections,
we shall try to clarify some consistency aspects of the search for a
stable internal manifold $B^d$.

\section{Dimensional reduction}

The spectral analysis of Kaluza-Klein theories, especially in the
internal sector, crucially depends on the consistency of the
dimensional reduction ansatz. The constraints \eq{2.15}, as all the
other constraints in our theory, should be preserved in time when
governed by the field equations \eq{2.10}. In addition, the new
constraints \eq{2.15} should be compatible with the ones already
present in the theory, such as \eq{2.11}. This means that no further
reduction of the number of degrees of freedom is expected. To see
how this works, we shall first rewrite the constraint equations
\eq{2.11} for our special case of the vacuum metric $M^4\times B^d$,
with $B^d$ an Einstein symmetric space. Thus, we find
\be
\left(\phi^{\,ln}\,G_{lM}\right)_{\!,\,n} +
\left(\ln\sqrt{-G}\right)_{\!,\,n}\,\phi^{\,ln}\,G_{lM} +
{1\over 2}\,\phi^{\,ln}_{\ \ ,\,M}\,G_{ln} = 0 \, ,        \lab{3.1}
\ee
with $G_{lm}=u_{lm}$, $G_{l\m}=B^k_{\m}\,u_{kl}$ and $G=gu$, as
follows from the decomposition \eq{2.12}. It is not difficult to
check the effect of dimensional reduction on the constraints
\eq{3.1}. When the ansatz \eq{2.15} is used, these become
\be
\left( \ln\sqrt{{u}\over{\phi}}\ \right)_{\!\!,\,n}
\left( \d^n_m + \vphi^n_m \right) -
{1\over 2}\, \vphi^n_{n\,,\,m} = 0 \, ,                  \lab{3.2}
\ee
\be
\left[\left( \ln\sqrt{{u}\over{\phi}}\ \right)_{\!\!,\,n}
\left( \d^n_m + \vphi^n_m \right) -
{1\over 2}\, \vphi^n_{n\,,\,m}\right]B^m_{\m} = 0 \, .
                                                         \lab{3.3}
\ee
As we can see, the first of the above equations implies the second,
and, consequently, it is only the constraint \eq{3.2} we shall be
occupied with in what follows. Remember that the variables
$\vphi_{mn}$ and $B^n_{\m}$ have the dimensionally reduced form
\eq{2.14}, with $K^m_a(y)$ the Killing vectors of the symmetric
space $B^d$, and $y$-independent related coefficients. To make use
of this fact, we shall continue our analysis by perturbative
methods. Expending the logarithm,
\[
\ln\sqrt{{u}\over{\phi}}\, =\, {1\over 2}\, \vphi^m_m -
{1\over 4}\, \vphi^n_m \,\vphi^m_n + \cdots \ ,
\]
one immediately finds the constraint \eq{3.2} to lack the linear
part. Explicitly,
\be
\left[ \,\left( K^n_a K_{cn}\right)\left( K^m_b K_{gm}\right)
f^{\ \ g}_{de} \,+\, a\leftrightarrow e \,\right]
\vphi^{ab}\vphi^{cd} = O(3) \, ,                          \lab{3.4}
\ee
where $f_{ab}^{\ \ c}$ are the structure constants of the isometry
group of $B^d$, as defined by
\be
K_a^m K^l_{b\,,\,m} - K_b^m K^l_{a\,,\,m} =
\a \, f_{ab}^{\ \ c} K^l_c \, .                             \lab{3.5}
\ee
The parameter $\a$ has the dimension of the inverse length, and is
introduced to make the structure constants $f_{ab}^{\ \ c}$
dimensionless. The expression in square brackets has a generic
nontrivial $y$-dependence. This means that, depending on the
particular $B^d$, the number of $y$-independent fields
$\vphi^{ab}(x)$ may strongly be reduced. For example, {\it in the
case of a two-sphere, one finds that \eq{3.4} forces all but the
trace component of $\vphi^{ab}$ to vanish}. This is why we have to
be careful with the interpretation of the results involving the
non-vanishing traceless components of $\vphi^{ab}$. An example of
the kind is the result of Ref. \cite{KKsigma} which states that
$B^d=S^2$ effective 4-dimensional theory has unstable traceless
modes in the scalar sector. Now, we see that the reliability of this
result is ruled out by the inconsistency of the dimensional
reduction used in its derivation. The trace mode alone, however, is
consistent with \eq{3.4}, and has {\it positive mass square}. In
fact, the scalar excitations of the form
\be
\vphi_{mn} \sim \vphi(x)\, \phi_{mn}(y)                    \lab{3.6}
\ee
satisfy the constraints \eq{3.2} in all orders and for any choice of
the internal manifold $B^d$. This is, however, not enough to ensure
the full consistency of the ansatz \eq{2.15} supplemented by
\eq{3.6}. Apart from the compatibility with the constraints of the
theory, one should also check if the ansatz is preserved in time
when governed by the full set of field equations. In linear
approximation, the equations of motion \eq{2.10} have the form
\bsubeq\lab{3.7}
\be
{\cal R}_{\mu\nu}+{1\over 2}\,\vphi_{,\,\mu\nu}
+{{2\l}\over{d+2}}\,\eta_{\m\n}\,\vphi
+{1\over 2}\,h_{\m\n ;\,l}^{\ \ \ \ \,l} = O(2) \, ,    \lab{3.7a}
\ee
\be
F^{\,\m\n}_{n\ ,\,\n}
+ (B^{\m}_{l\,;\,n}+B^{\m}_{n\,;\,l})^{;\,l}
- h^{\m\n}_{\ \ ,\,\n n}
+ {3\over 2}\,(h+\vphi)^{,\,\m}_{\ \ ,\,n} = O(2) \, ,  \lab{3.7b}
\ee
\bea
&&\Box\,\vphi_{mn} - 2\l \,\vphi_{mn} -
   2\,\RR_{kmln}\,\vphi^{kl} +
   \vphi_{mn ;\,l}^{\ \ \ \ \ l} +
   {{4\l}\over{d+2}}\,\phi_{mn}\,\vphi                  \nn \\
&&+\,2\,(\vphi + h)_{;\,mn} -
   (B^{\m}_{m\,;\,n}+B^{\m}_{n\,;\,m})_{,\,\m} =
   O(2) \, ,                                            \lab{3.7c}
\eea
\esubeq
where $F^l_{\m\n}\equiv B^l_{\m\n}-B^l_{\n\m}+O(2)$, $h\equiv
h^{\m}_{\m}\,$ and $\,\vphi\equiv \vphi^m_m\,$. In addition, the
linearized constraints \eq{2.11} read
\be
B^n_{\m\,;\,n} = O(2) \, , \qquad
(\vphi + h)_{;\,m} + 2\,\vphi^n_{m\,;\,n} = O(2)\, .     \lab{3.8}
\ee
Now, it is easy to see that the ansatz defined by \eq{2.15} and
\eq{3.6} brings the field equations \eq{3.7} into a form free of
$y$-dependent coefficients. Hence, the {\it linearized theory can
consistently be reduced to four dimensions}, and the resulting
effective theory turns out to be stable against small fluctuations
of the vacuum. The inclusion of the interaction terms, however,
spoils the nice character of this result. Although the linear part
of the field equations \eq{2.10} contains no $y$-dependent
coefficients, the higher order terms do. The restriction \eq{3.6}
then produces additional, unphysical constraints in the theory.
Therefore, the correct treatment of the generic internal excitations
must take care of their full $y$-dependence. This is, in particular,
true for the internal manifolds which have the form of coset spaces.
The full harmonic analysis of higher dimensional geometric
$\s$-models will be done elsewhere. Here, we turn our attention to
group manifolds.

When our internal manifold is chosen to have the structure of a semi
simple Lee group, the Killing vectors $K^l_a(y)\,$, $a=1,...,d\,$,
form a non-degenerate basis in $B^d$. It holds then,
\be
\phi^{\,mn} = K^m_a K^n_b \g^{ab} \ , \qquad
K^m_a K^n_b \phi_{mn} = \g_{ab} \, ,                    \lab{3.9}
\ee
where
\be
\g_{ab} \equiv
-{1\over 2}\,f_{ac}^{\ \ d}\,f_{bd}^{\ \ c}             \lab{3.10}
\ee
is the Cartan metric of the group, and $f_{ab}^{\ \ c}$ are the
corresponding structure constants. We shall use $\phi_{mn}$ and
$\g_{ab}$ to raise and lower the world and group indices,
respectively. If we now apply the ansatz \eq{2.15} to our constraint
equations \eq{3.1}, and analyze the resulting expression \eq{3.4},
we shall find that it is identically satisfied. Therefore, no
further reduction of the number of degrees of freedom occurs. We do
not need the additional constraint \eq{3.6}, and this holds true in
all orders, as is seen  by the inspection of the non-perturbative
expression \eq{3.2}. Similarly, one can analyze the very equations
of motion. By projecting them on the Killing basis, we shall obtain
the equations for our $y$-independent variables $g_{\m\n}(x)\,$,
$A^a_{\m}(x)$ and $\vphi^{ab}(x)\,$, with the coefficients
consisting of $d$-scalar combinations of the Killing vectors and
their covariant derivatives. Now, we notice that, for group
manifolds, it holds
\be
K_{am\,;\,n} = {{\a}\over 2}\,f_{abc}\,K^b_m K^c_n \, . \lab{3.11}
\ee
Therefore, the Killing vector derivatives are fully expressed in
terms of the Killing vectors themselves. Being $d$-scalars, the
coefficients of our field equations boil down to completely
contracted products of the Killing vectors. It follows from \eq{3.9}
then that these coefficients contain only constant group tensors
$\g_{ab}$ and $f_{ab}^{\ \ c}$. The resulting field equations have
no $y$-dependent coefficients, and we conclude that {\it dimensional
reduction of group manifolds is a consistent procedure}.

We now want to analyze the mass spectrum of the corresponding
effective 4-dimensional theory. When $B^d=$ a group manifold, the
linearized equations of motion \eq{3.7} become
\bsubeq\lab{3.12}
\be
{\cal R}_{\mu\nu}+{1\over 2}\,\vphi_{,\,\mu\nu}
+{{2\l}\over{d+2}}\,\eta_{\m\n}\,\vphi = O(2) \, ,     \lab{3.12a}
\ee
\be
F^{\mu\nu}_{a\ ,\,\nu} = O(2)  \, ,                    \lab{3.12b}
\ee
\be
\Box\,\vphi_{ab} +
\mu_{abcd}\,\vphi^{cd} = O(2) \, .                     \lab{3.12c}
\ee
\esubeq
with the mass matrix $\mu_{abcd}$ given by
\be
\mu_{abcd} \equiv
2\l\left({2\over{2+d}}\,\g_{ab}\g_{cd} -
f_{ac}^{\ \ e}f_{bde}\right) .                       \lab{3.13}
\ee
The parameter $\l$ is related to the coupling constant $\a$ by
\[
\l \equiv -{{\a^2}\over 2} \, ,
\]
as follows from the well known form of the curvature tensor for
group manifolds:
\[
\RR_{abcd} = -{{\a^2}\over 4}\,f_{ab}^{\ \ e}f_{cde}\, .
\]
The first two equations \eq{3.12} describe the well known Einstein
and Yang-Mills sectors of the theory, and will not be examined
further. The stability of the Klein-Gordon sector, however,
crucially depends on the signature of the mass matrix $\mu_{abcd}$,
and, thus, on the particular group considered. One immediately sees,
for example, that 3-dimensional groups necessarily carry negative
modes. Indeed, owing to $f_{abc}\sim \ve_{abc}$ and $\g_{ab}\sim
\d_{ab}$, the traceless modes of $\vphi^{ab}$ are easily seen to
have negative mass terms. In particular, the groups $SU(2)$ and
$SO(3)$ define unstable theories. Similarly, the group $SO(4)$,
being locally isomorphic to $SU(2)\times SU(2)$, gives an equally
unattractive result. In the next section, we shall analyze the
spectrum of the generic $SO(n)$ groups.

\section{SO(n) group manifolds}

The $SO(n)$ group generators are commonly denoted by $M_{ij}\equiv
-M_{ji}\,$, $i,j = 1,...,n\,$, and are subject to the commutation
relations
\be
\left[\,M_{ij}\ ,\, M_{kl}\,\right] =
{{\a}\over 2}\,f_{ijkl}^{\ \ \ \ mn}\,M_{mn}\, ,         \lab{4.1}
\ee
with
\be
f_{ijkl}^{\ \ \ \ mn} \equiv
2\left(\d^m_{[i}\,\d^{\vphantom{m}}_{j][k}\,\d^n_{l]}
\ -\ m \leftrightarrow n \right) \, .                    \lab{4.2}
\ee
In this section, the indices $i,j,...$ are taken to run from 1 to
$n$, and should not be confused with internal indices of the
preceding sections. The group indices $a,b,...$ are seen as
antisymmetric pairs of indices $i,j,...\,$. In accordance with
\eq{3.10}, the Cartan metric of the group is found to be
\be
\g_{ijkl} = {{n-2}\over 2}\,\d_{i[k}\,\d_{jl]} \, .      \lab{4.3}
\ee
Our task in this section is to analyze the mass spectrum of the
scalar sector of the theory, as given by \eq{3.12c}. To this end, we
shall decompose the variables $\vphi_{ab}\equiv \vphi_{ijkl}$ ($\,a
\rightarrow ij\,$, $b \rightarrow kl\,$) into irreducible components
which diagonalize the mass matrix \eq{3.13}. Thus, we obtain
\be
\vphi_{ijkl} = \tilde{\vphi}_{ijkl} + {4\over {n-2}}\,
\d_{[i[k}\,\tilde{\vphi}_{j]l]} + {2\over {n(n-1)}}\,
\d_{[ik}\,\d_{j]l}\,\tilde{\vphi} \, ,                   \lab{4.4}
\ee
where $\tilde{\vphi}_{ijkl}$ and $\tilde{\vphi}_{ij} =
\tilde{\vphi}_{ji}$ are traceless, and
\be
\tilde{\vphi}_{ij} \equiv
\vphi_{ij}-{1\over n}\,\d_{ij}\,\tilde{\vphi} \,,\qquad
\vphi_{ij} \equiv \d^{kl}\,\vphi_{ikjl}\,,\qquad
\tilde{\vphi} \equiv \d^{ij}\,\vphi_{ij} \, .            \lab{4.5}
\ee
The traceless component $\tilde{\vphi}_{ijkl}$ is still reducible,
and can further be decomposed into totally antisymmetric part and
the rest:
\be
\tilde{\vphi}_{ijkl} = A_{ijkl} +
{2\over 3}\left(S_{ikjl} - S_{iljk}\right)\,,            \lab{4.6}
\ee
where
\be
A_{ijkl} \equiv {1\over 3}\,\left(
\tilde{\vphi}_{ijkl} + \tilde{\vphi}_{iklj} +
\tilde{\vphi}_{iljk}\right) \, ,        \qquad
S_{ijkl} \equiv {1\over 2}\,\left(
\tilde{\vphi}_{ikjl}+\tilde{\vphi}_{iljk}\right)\, .     \lab{4.7}
\ee
The quantity $A_{ijkl}$ is totally antisymmetric, while $S_{ijkl}$,
in addition to being traceless and symmetric with respect to
$i \leftrightarrow j\,$, $k \leftrightarrow l$ and $ij
\leftrightarrow kl\,$, satisfies the cyclic identity
\[
S_{ijkl}+S_{iklj}+S_{iljk} = 0 \, .
\]
The components $\tilde{\vphi}\,$, $\tilde{\vphi}_{ij}\,$, $A_{ijkl}$
and $S_{ijkl}$ are irreducible components of $\vphi_{ijkl}\,$. The
dimensions of the corresponding irreducible subspaces ($n>2$) are:
\[
1 \, , \qquad {n+1 \choose 2} - 1 \, , \qquad
{n \choose 4} \qquad \mbox{and} \qquad
\frac{n-3}{2} \, {n+2 \choose 3} \, ,
\]
respectively. Their sum gives the total number of $\frac 12 \,
d(d+1)$ independent components $\vphi_{ab}\,$, where $d=\frac 12 \,
n(n-1)$ for $SO(n)$ group manifolds.

Now, we shall apply the decomposition \eq{4.4}, \eq{4.6} to the
Klein-Gordon sector of the field equations \eq{3.12}. We have
already seen that the trace component $\vphi\equiv \vphi^a_a$ has
positive mass square independently of the choice of $B^d$. Indeed,
our $\tilde{\vphi}$, being proportional to $\vphi$, is explicitly
found to satisfy
\be
\left(\Box - {{8\l}\over{d+2}}\right)\tilde{\vphi}
= O(2) \, ,                                              \lab{4.8}
\ee
with $\l\equiv -{1\over 2}\,\a^2$ ensuring the correct sign of the
mass term. A similar result is obtained for the totally
antisymmetric irreducible component $A_{ijkl}\,$. The cumbersome,
but otherwise simple, calculations lead to
\be
\left(\Box - {{8\l}\over{n-2}}\right)A_{ijkl}=O(2) \, .  \lab{4.9}
\ee
The content of these equations is nontrivial only for $n>3$ because
the totally antisymmetric $A_{ijkl}$ otherwise vanishes. Therefore,
the corresponding mass term is positive in all nontrivial cases.

This is not quite so in the case of $\tilde\vphi_{ij}$ components.
Evaluating the mass matrix \eq{3.13} for the $SO(n)$ structure
constants \eq{4.2}, and diagonalizing it by \eq{4.4}, one finds
\be
\left( \Box - 2\l \,{{n-4}\over{n-2}}\, \right)
\tilde\vphi_{ij} = O(2) \, .                            \lab{4.10}
\ee
We see that $SO(3)$ group manifolds contain 5 unstable modes in the
Klein-Gordon sector, which restrains us to consider only $n\geq 4$
cases. The stable massive modes, having masses of the order of the
Planck mass, do not appear in the effective low energy theory. An
exception is the $SO(4)$ theory which accommodates 9 zero-mass modes.
In any case, the irreducible components $\tilde{\vphi}\,$,
$\tilde{\vphi}_{ij}$ and $A_{ijkl}$ are all stable if $n\geq 4$.

Finally, the irreducible component $S_{ijkl}$ is found to satisfy
the Klein-Gordon equation of the form
\be
\left(\Box +{{4\l}\over{n-2}}\right)S_{ijkl}=O(2) \, .  \lab{4.11}
\ee
As we can see, the mass term of \eq{4.11} is negative in all
nontrivial cases. The $SO(3)$ theory, which does not accommodate
either $A_{ijkl}$ or $S_{ijkl}$ modes, is ruled out by the presence
of unstable $\tilde{\vphi}_{ij}$ components in \eq{4.10}. Therefore,
{\it neither of $SO(n)$ group manifolds can be used as a stable
internal space} of our Kaluza-Klein geometric $\s$-model.

One could try to improve this situation by imposing additional
constraints (such as $S_{ijkl}=0$) to the theory. However, losing
kinetic terms for some field components can lead to the appearance
of new, unphysical constraints stemming from the interaction part of
field equations. Moreover, even if we find a consistent set of
constraints to define our dimensional reduction, we cannot be sure
how infinitesimal perturbations of the ansatz itself affect the
whole scheme. In other words, {\it not only the consistency, but
also the stability of the dimensional reduction prescription is
needed}. To be more specific, if the constraints \eq{2.15} are
perturbed by adding small $y$-dependent terms, these may evolve to
considerably change the initial ansatz. To prevent this, one must
take into consideration the full $y$-dependence of the theory. Only
the complete $(4+d)$-dimensional stability of the vacuum can lead to
a correctly reduced effective theory. Once the masses of all higher
dimensional excitations are proven positive, the effective theory is
obtained by discarding heavy mass modes. The influence of higher
modes in the harmonic decomposition of fields in Kaluza-Klein
$\s$-models will be considered elsewhere.

\section{Concluding remarks}

The analysis of the preceding sections has mainly been devoted to
the consistency of dimensional reduction of higher dimensional
geometric $\s$-models. The motivation came from the failure of
Ref. \cite{KKsigma} to provide an example of a stable $M^4
\times B^d$ vacuum of the theory. In particular, the excitations of
the internal manifold $S^2$ were shown to possess unstable modes. In
this paper, the search for a stable $B^d$ has been required to
respect a consistent scheme of dimensional reduction. With this in
mind, the constraints of the theory were analyzed in detail in
Section III. When dimensionally reduced, these constraints have been
shown to confine the number of independent excitations of a generic
internal manifold $B^d$. In particular, the consistently reduced
$M^4 \times S^2$ theory turned out not to accommodate the traceless
scalar excitations of the vacuum. The trace mode alone, however, has
been proven stable independently of the specific $B^d$ used. The
restriction of the internal excitations to their trace mode has also
been shown to respect the full set of linearized field equations.
Unfortunately, the inclusion of higher order terms leads to the
appearance of new, unphysical constraints in the theory, which
brings us to the conclusion that the correct treatment of generic
vacuum excitations must take care of their full $y$-dependence. In
the last part of Section III, we have analyzed the role of group
manifolds in defining the internal spaces of our Kaluza-Klein
geometric $\s$-models. It has been demonstrated how the full
interactive theory can be consistently reduced to 4 dimensions.

In Section IV, we have examined the example of a generic $SO(n)$
group. Decomposing the scalar fields $\vphi_{ab}$ into their
irreducible components, the mass matrix of the Klein-Gordon sector
of the theory has successfully been diagonalized. It turned out,
however, that no $SO(n)$ group led to a positive definite mass
matrix.

The solution of the problem could hardly be found in additional
restrictions of the theory (such as rejecting the unstable modes),
since losing some of the modes usually leads to the appearance of
new, unphysical constraints. It has been argued that it is not
enough to have a consistent set of constraints which define
dimensional reduction, but also that the effective theory should be
stable against small perturbations of the ansatz itself. In this
respect, having found a group manifold leading to a stable
4-dimensional theory, one is still left with the task to check the
influence of higher harmonics. Only the full higher dimensional
stability of the vacuum can lead to a correctly reduced effective
theory. The harmonic analysis of internal manifolds of higher
dimensional geometric $\s$-models, especially $d$-spheres, will be
considered elsewhere.



\begin{thebibliography}{99} 
%
\bibitem{sigma}
M. Vasili\'c, Class. Quantum Grav. {\bf 15}, 29 (1998).
%
\bibitem{Finkelstein}
D. Finkelstein and J. Rubinstein, J. Math. Phys. {\bf 9},
1762 (1968).
%
\bibitem{Misner}
D. Finkelstein and C. W. Misner, Ann.Phys. NY {\bf 6}, 230 (1959).
%
\bibitem{Olive}
P. Goddard and D. I. Olive, Rep. Prog. Phys. {\bf 41}, 1357 (1978).
%
\bibitem{Ringwood}
G. A. Ringwood and L. M. Woodward, Phys. Rev. Lett. {\bf 47},
625 (1981).
%
\bibitem{KKsigma}
M. Vasili\'c, Phys. Rev. D {\bf 60}, 25003 (1999).
%
\bibitem{Omero}
C. Omero and R. Percacci, Nucl. Phys. B {\bf 165}, 351 (1980).
%
\bibitem{Gell-Mann}
M. Gell-Mann and B. Zwiebach, Phys. Lett. {\bf 141B}, 333 (1984).
%
\bibitem{Salam}
A. Salam and J. Strathdee, Ann. Phys. NY {\bf 141}, 316 (1982).
%
\end{thebibliography}
\end{document}